\documentclass[nofootinbib,prd,preprintnumbers,showpacs,superscriptaddress,twocolumn]{revtex4}

\usepackage{epsfig,amssymb}
\usepackage{url}

\def\beq{\begin{equation}}
\def\eeq{\end{equation}}
\def\bea{\begin{eqnarray}}
\def\eea{\end{eqnarray}}
\def\d{{\mathrm{d}}}

\newfont{\cursive}{pzcmi at 9pt}
\def\~t{\tilde{t}}

\begin{document}

\newcommand*{\AEI}{Max-Planck-Institut f\"ur
    Gravitationsphysik, Albert-Einstein-Institut, Am M\"uhlenberg 1,
    D-14476 Golm, Germany}\affiliation{\AEI}

\newcommand*{\CCT}{Center for Computation \& Technology, 216
    Johnston Hall, Louisiana State University, Baton Rouge, LA 70803,
    USA}\affiliation{\CCT}
\newcommand*{\LSU}{Department of Physics \& Astronomy, 202
    Nicholson Hall, Louisiana State University, Baton Rouge, LA 70803,
    USA}\affiliation{\LSU}

  \title{The slicing dependence of non-spherically symmetric
    quasi-local horizons in Vaidya Spacetimes}

\author{Alex B. Nielsen}\email{alex.nielsen@aei.mpg.de}\affiliation{\AEI}
\author{Michael Jasiulek}\email{michael.jasiulek@aei.mpg.de}\affiliation{\AEI}
\author{Badri Krishnan}\email{badri.krishnan@aei.mpg.de}\affiliation{\AEI}
\author{Erik Schnetter}\email{schnetter@cct.lsu.edu}\affiliation{\CCT}\affiliation{\LSU}

\begin{abstract}

  It is well known that quasi-local black hole horizons depend on the
  choice of a time coordinate in a spacetime.  This has implications
  for notions such as the surface of the black hole and also on
  quasi-local physical quantities such as horizon measures of mass and
  angular momentum.  In this paper, we compare different horizons on
  non-spherically symmetric slicings of Vaidya spacetimes.  The
  spacetimes we investigate include both accreting and evaporating
  black holes. For some simple choices of the Vaidya mass function
  function corresponding to collapse of a hollow shell, we compare the
  area for the numerically found axisymmetric trapping horizons with
  the area of the spherically symmetric trapping horizon and event
  horizon. We find that as expected, both the location and area are
  dependent on the choice of foliation. However, the area variation is
  not large, of order $0.035\%$ for a slowly evolving horizon with
  $\dot{m}=0.02$.  We also calculate analytically the difference in
  area between the spherically symmetric quasi-local horizon and event
  horizon for a slowly accreting black hole. We find that the
  difference can be many orders of magnitude larger than the Planck
  area for sufficiently large black holes.


\end{abstract}


\maketitle


\section{Introduction}

An important feature of black hole spacetimes is the existence of
trapped and marginally trapped surfaces.  These surfaces are a key
element in the Penrose singularity theorem and are frequently used in
numerical relativity to indicate the existence of a black hole. Recent
work has introduced related concepts such as trapping horizons and
dynamical horizons both of which, ignoring certain technical
conditions, can be viewed 3-surfaces obtained by time-evolutions of
marginally trapped surfaces.  It has been shown that these surfaces
have many properties similar to event horizons, in particular their
thermodynamic properties
\cite{Ashtekar:2004cn,Booth:2005qc,Nielsen:2008cr}. Trapping and
dynamical horizons can also be used to extract data about the black
hole, such as its mass and angular momentum, directly from the strong
gravity region without needing to rely on the asymptotic behaviour or
exact isometries \cite{Ashtekar:2004cn}.

However, it has been known for a long time that the location and
existence of trapped surfaces depends on the choice of spacetime
foliation, i.e. on the choice of the time coordinate, and this is the
main reason why event horizons have often been preferred over trapping
and dynamical horizons as a definition of the surface of a black
hole. In particular this means that any given, suitably general,
spacetime that contains a trapping horizon is likely to contain, in
principle, an infinite number of trapping and dynamical horizons that
intersect but are distinct \cite{Ashtekar:2005ez}.  The key results
for quasi-local horizons, for example the laws of black hole mechanics
and the energy and angular momentum balance laws
\cite{Ashtekar:2003hk}, are sufficiently general and are applicable to
all of these different horizons.  Nevertheless, if we wish to assign
say mass and spin to the black hole quasi-locally, the time evolution of
these quantities will depend on which horizon we choose to use.  Thus,
to use quasi-local horizons to extract meaningful data from
black hole spacetimes it is important to investigate the location of
these distinct horizons and how this affects their related properties
such as mass, angular momentum, linear momentum etc.

This issue is potentially relevant for gravitational wave astronomy
when one seeks to infer the physical parameters of the source from the
observed gravitational wave signal.  For example, in modeling binary
black hole coalescence it becomes important to combine results from
post-Newtonian theory and numerical relativity.  Numerical relativity
can in principle evolve a binary black hole system starting from the
inspiral phase all the way through the coalescence (see
e.g. \cite{Hinder:2010vn} for a recent review).  However, this becomes
computationally impossible if we start the black holes very far apart,
and if we wish to cover a large parameter space.  Post-Newtonian theory
on the other hand treats the black holes as point particles endowed
with a mass and spin, and solves the field equations in powers of
$v/c$ with $v$ being the orbital velocity of the black holes.  This
approximation works best when the black holes are far apart and breaks
down near the coalescence.  Thus, in order to fully solve the binary
black hole problem it becomes essential to combine the two frameworks.
The literature on this topic is large and a full discussion is beyond
the scope of this paper; see e.g.
\cite{Buonanno:2006ui,Baker:2006ha,Campanelli:2008nk,Boyle:2007ft,Hannam:2007ik,Hinder:2008kv,Santamaria:2010yb}.
For our purposes, we note that the two frameworks treat black holes
very differently; post-Newtonian theory treats the black holes as
point particles while numerical relativity deals with the horizons
non-perturbatively.  It turns out that using different flavors of
Post-Newtonian approximants leads to biases in the values of physical
parameters \cite{Buonanno:2009zt} (as infered from the gravitational
waveforms).  Numerical relativity on the other hand deals with black
hole horizons, and if we decide to use quasi-local horizon measures
for calculating physical quantities, the values of the black hole
parameters will be affected by the choice of foliation.  This effect
is absent when the black holes are isolated; thus we expect it to be
negligible in the inspiral phase and to be larger in the dynamical
coalescence phase.  In comparing the results from the two frameworks,
it is useful then to quantify this source of uncertainty and to show
that errors in the physical parameters are smaller than the biases
between, say, different post-Newtonian approximants.

Another area where the location of horizons is important is in black
hole thermodynamics. If the area of a black hole horizon is to play a
role analogous to entropy via the Bekenstein-Hawking relation it is
important to know which area this relation should be applied to in
which situations. For stationary black holes, cross-sections of the
event horizon are marginally trapped surfaces and thus the different
notions of black hole horizons (i.e. event horizons and trapping
horizons) coincide.  However, in a dynamical spacetime the event
horizon will not coincide with the location of any of the trapping
horizons.  In general they will have different areas on a given
spacelike hypersurface \cite{Nielsen:2010gm,Booth:2010eu}. Furthermore, although the location of the
event horizon does not depend on the choice of spacetime foliation,
when the event horizon is growing or shrinking, its area will depend
on the choice of foliation.


Non-stationary spacetimes with dynamical black holes are complicated,
and only a few exact dynamical black hole solutions are known. One
class of solutions is the Vaidya solutions, which describe the
evolution of a spherically symmetric, radially moving, pressureless
null fluid with a freely specifiable mass function $m(v)$. This is a
useful and non-trivial toy-model for a dynamical black hole.  These
spacetimes can be used to qualitatively model the physical accretion
of matter by a black hole. Some authors
\cite{Hiscock:1980ze,Hayward:2005gi} have even suggested using the
Vaidya spacetimes to simulate the decrease in area of a black hole due
to Hawking radiation, by allowing the infalling matter to be negative.

The uniqueness of dynamical horizons in general spacetimes was
investigated in \cite{Ashtekar:2005ez} where it was shown that the
foliation of a given dynamical horizon by marginally trapped surfaces
is unique, and that it is not possible to foliate a region with
dynamical horizons since they intersect. In the Vaidya solution it has
already been shown explicitly that marginally outer trapped surfaces
(MOTS) can be found at various different locations depending on the
slicing \cite{Schnetter:2005ea}, although this reference did not
investigate how the parameters of the black hole, such as area, mass
and angular momentum vary with slicing and did not consider the case
of timelike trapping horizons.  Ben-Dov \cite{BenDov:2006vw} showed
how to find marginally outer trapped surfaces arbitrarily close to the
event horizon.  It was also shown that there are flat regions inside the
event horizon (the region contained within the shell) where no trapped
surface passes.  Bengtsson and Senovilla \cite{Bengtsson:2008jr} analytically
constructed trapped surfaces that pass through the flat region. This
is not a contradiction; while there are some flat regions where no
trapped surface can be located, there are trapped surfaces in other
parts of the flat region.
In spherically symmetric spacetimes spherically symmetric marginally
trapped surfaces are easy to find and in the case of Vaidya are given
by the condition $r=2m(v)$, where $r$ is the areal coordinate
($r=\sqrt{A/16\pi}$ with $A$ being the area of the MOTS) of the
surfaces of spherical isometry. It is also easy to show that such
surfaces foliate a trapping horizon which we will refer to as the
spherically symmetric trapping horizon. Due to results in
\cite{Ashtekar:2005ez}, any other dynamical horizon will lie partially
outside the spherically symmetric dynamical horizon somewhere. For
example, in \cite{Aman:2009bv} analytic solutions were presented where
closed trapped surfaces extend into the region between the spherically
symmetric trapping horizon and the event horizon. The surfaces we find
here do the same.

In this article we will look at the variation of black hole parameters
for horizons located on various different slicings of the same
spacetime. We will examine several different spherically symmetric
mass functions; a linear mass function designed to see the behaviour
for slowing evolving black holes and a tanh log mass function,
designed to see the behaviour in a short collapse of a shell to a
black hole. We compare marginally outer trapped surfaces found on
non-spherically symmetric hypersurfaces with those found on the
spherically symmetric slicings. We are able to locate the spherically
symmetric horizons analytically but the non-spherically symmetric
horizons are located numerically using the AHFinderDirect thorn
\cite{Thornburg:1996, Thornburg:2003sf} of the Cactus framework
\cite{Goodale:2002, cactus}. A number of useful analytic relations for
the Vaidya spacetime are given in the appendix. Background detail on
the properties of trapping horizons and their thermodynamics can be
found in \cite{Ashtekar:2004cn} and \cite{Nielsen:2008cr}.

\section{The Vaidya metric}


To examine the slicing dependence of quasi-local horizons we need to
consider dynamical spacetimes.  The Vaidya solutions have a number of
nice properties that make them popular for investigations of this type
\cite{Schnetter:2005ea,BenDov:2006vw,Bengtsson:2008jr}. The Vaidya
spacetimes \cite{Vaidya:1951zz} are a class of spherically symmetric, non-vacuum spacetimes
with line element in advanced null Eddington-Finkelstein coordinates
$(v,r,\theta,\phi)$:
\beq \d s^2 = - \left( 1-\frac{2m(v)}{r} \right) \d v^2 + 2\d v\d r + r^{2}\d\theta^2 + r^{2}\sin^{2}\theta\d\phi^2 .\eeq
The function $m(v)$ is a a freely specifiable mass function; it
coincides with the Misner-Sharp mass \cite{MisnerSharp} in this
case. In advanced Eddington-Finkelstein coordinates the metric is well
defined across future horizons. The Vaidya metric solves the Einstein
equations with an energy-momentum tensor of the form
\beq T_{ab} = E n_{a} n_{b}, \eeq
where $n^{a}$ is an ingoing radial null direction and $E$ is a function that depends on the normalization of $n^{a}$. For example, for the case where $n_{a} = -\partial_{a}v$ then we have
\beq \label{NEC} T_{ab}\ell^{a}\ell^{b}=\frac{\dot{m}}{4\pi r^{2}}, \eeq
when $n^{a}\ell_{a} = -1$ and defining $\dot{m} = \frac{\d m(v)}{\d v}$.
The Vaidya solution can be interpreted as describing the radial
collapse of pressureless null dust (i.e.\ infalling radiation).  From
the appendix we can see that, in a spherically symmetric foliation,
trapping horizons will occur at $r=2m(v)$ provided we keep $m(v)>0$.

Here we will study various mass functions for $m(v)$. The simplest is a linear mass function of the form
\beq \label{linearmass} m(v) = m_{0} + \dot{m}v .\eeq
The linear mass function is suited to situations where the black hole
is accreting matter at a constant rate (for a finite time duration to
ensure that the mass stays finite). While the assumption of spherical
symmetry is somewhat artificial, real astrophysical black holes do
have very small mass accretion rates whether from the accretion of
surrounding stars and gas or purely from the accretion of photons from
the cosmic microwave background. In the Vaidya solutions the
Misner-Sharp mass on each ingoing constant $v$ surface is a
constant. This reflects the fact that the mass is flowing inwards at
the speed of light and the mass contained within a shell of radius $r$
is constant as the radius decreases. Although the spacetime is
dynamical we can still define observers who remain at a fixed areal
radius $r$ and fixed $\theta$ and $\phi$. In terms of the proper time
$\tau$ of such observers we have
\beq \frac{\d v}{\d \tau} = \frac{1}{\sqrt{1-2m(v)/r}} .\eeq
At large distances from the black hole $r \gg m(v)$ we can use this to relate directly the mass flux in terms of the null coordinate $v$ to the mass flux as seen by constant $r,\theta,\phi$ observers who would be static observers in an exactly static spacetime,
\beq \frac{\d m(v)}{\d v} \sim \frac{\d m(v)}{\d \tau} .\eeq
We can give a very rough indication of the order of magnitude for $\dot{m}$ that might be expected for certain astrophysical cases. The Eddington rate is used to estimate the maximal rate at which infalling matter can be supported by its own radiation pressure and large, luminous black holes are typically found with luminosities between $10$ and $100\%$ of the Eddington limit \cite{Kollmeier:2005cw}. For a black hole accreting at a tenth of the Eddington limit the dimensionless accretion rate is approximately \cite{Nielsen:2010gm}
\beq \dot{m} \simeq 10^{-22}\left( \frac{M}{M_{\odot}} \right) .\eeq
This is $\sim 10^{38}$ ergs per second for a solar mass black hole.
The matter falling into the black hole is ten times the energy being
emitted as light.
Since this accretion is usually associated with a disk though it will
not be spherically symmetric.  For a black hole accreting purely from
the Cosmic Microwave Background, which is assumed to be isotropic, we
have from the Stefan-Boltzmann law, approximately
\beq \dot{m} \simeq 10^{-50}\left(\frac{T}{T_{_{3K}}}\right)^{4}\left(\frac{M}{M_{\odot}}\right)^{2} .\eeq
For a black hole whose dynamics are dominated by evaporation through Hawking radiation we have
\beq \dot{m} \simeq -10^{-81}\left(\frac{M_{\odot}}{M}\right)^{2} .\eeq
For numerical purposes we will investigate evolutions with mass rates much higher than astrophysical rates, typically $|\dot{m}| \sim 0.01$. To the extent that $2\sqrt{\dot{m}} \ll 1$, the spherically symmetric trapping horizon will still be a slowly evolving horizon in the sense of \cite{Booth:2006bn}. We will also look at mass functions of the $\tanh \; \log$ form, with $m(v)=0$ for $v<0$ and for $v>0$
\beq \label{tanhlogmass} m(v) = \frac{m_{0}}{2}\left(1 + \tanh\left(\log\left( \frac{v}{T}\right)\right)\right) = \frac{m_{0}v^2}{v^2 + T^2}\eeq
%
%
%
The first derivative of this tanh log mass function vanishes at $v=0$. Therefore the metric and its first derivative are contiunous at $v=0$. For $T$ and $m_{0}$ greater than zero, this mass function models the collapse of a hollow spherical shell of matter $m(v=0)=0$ that asymptotically settles down to an isolated black hole of mass $m_{0}$ for $v\rightarrow\infty$. These types of mass functions model situations where the black hole grows initially quite rapidly and then asymptotically settles down to its final static state. The mass function reaches half its asymptotic value when $v=T$.

The maximum of $\dot{m}$ for these tanh log functions occurs at $v=T/\sqrt{3}$ and takes a value $\sim m_{0}/T$. Therefore these solutions will not be slowly evolving in the sense of \cite{Booth:2006bn} for $m_{0}$ and $T$ of similar sizes.

\subsection{Location of spherically symmetric horizons}
The trapping horizons are three-dimensional surfaces $H$, foliated by
closed spacelike two surfaces for which the future directed null
normals $\ell^{a}$ and $n^{a}$ satisfy
\beq \theta_{(\ell)} =  0\,, \qquad
\theta_{(n)}  <  0\,, \qquad
{\cal{L}}_{n}\theta_{(\ell)} <  0\,. \eeq
Here $\theta_{(\ell)}$ and $\theta_{(n)}$ are the expansions of
$\ell^a$ and $n^a$ respectively, and $\cal{L}_{n}$ is the Lie derivative
along $n^a$.  Dynamical horizons are also three-dimensional surfaces
foliated, as above, by spheres with $\theta_{(\ell)} = 0$,
$\theta_{(n)} < 0$. However, ${\cal{L}}_{n}\theta_{(\ell)} < 0$ is
replaced by the requirement that $H$ be spacelike.  For the
spherically symmetric horizons in Vaidya, these notions coincide
\cite{Ashtekar:2004cn}.

For spherically symmetric slicings, the null normals associated with
the trapping horizon will be radial null vectors and the location of
the trapping horizon is just given by
\beq r=2m(v) \label{r_is_two_m} \,.\eeq
This is a spacelike surface for $\dot{m} > 0$, a null surface for $\dot{m}=0$ and a timelike surface for $\dot{m} < 0$. For the linear mass function, in terms of the timelike coordinate $t=v-r$, the horizons will be located at
\beq r = \frac{2m_{0} + 2\dot{m}t}{1-2\dot{m}} \label{r_is_not_two_m}. \eeq
For the $\tanh\; \log$ mass function the horizons are located at the solution of the cubic
\beq r^{3} + \left(2t-2m_{0}\right)r^{2} + \left( t^{2}+T^{2}-4m_{0}t\right)r - 2m_{0}t^{2} = 0 .\eeq
The equation for the horizon $r=2m(v)$ has a single unique solution on each surface of constant $v$. But it can have multiple solutions, corresponding to multiple horizons, on surfaces of constant $t$. For $m_{0} > 0$ this cubic function always has at least one positive real root and guarantees that there will always be at least one horizon. In fact, it can be shown that there will only be a single horizon for $t>2m_{0}-\sqrt{4m_{0}^2-T^2}$. For small values of $t$ there are multiple horizons and this will also be the case for other mass functions.

We will use numerical methods to solve the trapping horizon equations for non-spherically symmetric slicings and postpone discussion of them until the next section. Due to results in \cite{Ashtekar:2005ez} we expect that the non-spherically symmetric trapping horizons will intersect the spherically symmetric ones.

The event horizons are defined as the past causal boundary of future null infinity and are generated by null geodesics that fail to reach infinity. The event horizon is always a null surface since it is a causal boundary. In the Vaidya spacetimes it is generated by radial outgoing null vectors that satisfy
\beq \frac{\d r}{\d v} = \frac{1}{2}\left(1-\frac{2m(v)}{r}\right) .\label{ehlocation} \eeq
This first order ordinary differential equation generates the path of all outgoing radial null geodesics. In order to give the location of the event horizon it requires a boundary condition that corresponds to the known location of the event horizon at some particular point. In practice this is usually given by the position of the event horizon at some future point, either when the black hole evaporates entirely or settles down to a stationary state. If the black hole at some point settles down to a Schwarzschild black hole with no further matter accreting, then the event horizon can be located by tracing back the null rays from the future Schwarzschild radius. However, in the situation where the black hole is accreting matter at a steady rate and is a suitably long way from changing to a different state one can find the approximate location of the event horizon by imposing the condition
\beq \frac{\d^{2}r}{\d v^{2}} = 0 .\eeq
This just reflects the fact that the event horizon is growing at a steady rate \cite{Nielsen:2010gm}. In this case, equation (\ref{ehlocation}) has the general solution
\beq r = \frac{m(v)}{4\dot{m}}\left(1-\sqrt{1-16\dot{m}}\right) .\eeq
For $\dot{m} \ll 1$ this gives
\beq \label{rehsmallmdot} r = 2m(v)\left(1 + 4\dot{m} + {\mathcal{O}}(\dot{m}^2)\right) \eeq
and thus we expect that the event horizon will be outside the spherically symmetric trapping horizon for $\dot{m} > 0$ but inside for $\dot{m} < 0$. For a solar-mass black hole accreting at a tenth of the Eddington rate the difference in areas between the event horizon and the spherically symmetric trapping horizon will be around $10^{56}$ in units of Planck area, while for a supermassive black hole of mass $10^{8}$ solar masses, accreting purely form the CMB, the difference in areas will be around $10^{60}$ in Planck units \cite{Nielsen:2010gm}. In terms of the coordinate $t=v-r$ for the linear mass function the event horizon has radial coordinate
\beq r \sim \frac{2m+2\dot{m}t}{1-2\dot{m}} + \frac{8m\dot{m}}{1-2\dot{m}}.\eeq
This is just the location of the spherically symmetric trapping horizon with a constant offset of $8m\dot{m}$ provided $\dot{m} \ll 1$. In this approximation the generators of both the trapping horizon and the event horizon have the same components but the norm of the generators is $4\dot{m}$ for the trapping horizon and zero for the event horizon. 

\subsection{An axisymmetric spacetime slicing}

We consider a simple axisymmetric slicing of the form
\beq\label{eq:tvralphaz} \bar{t} = v-r - \alpha z ,\eeq
where $z = r \cos\theta$ and $\alpha$ is a parameter that determines how far away from spherically symmetry the constant $\bar{t}$ surface is. We reserve the symbol $t$ for spherically symmetric surfaces where $\alpha = 0$, which just gives the usual Eddington-Finkelstein time coordinate (although not the Schwarzschild time coordinate). Hypersurfaces of constant $\bar{t}$ are always timelike for $ |\alpha |<1$ since the normal to a given hypersurface, $\bar{t}^{a}$, has norm
\beq \bar{t}^{a}\bar{t}_{a} = \alpha^2 - 1 - \frac{2m(v)}{r}\left( 1 + \alpha \cos\theta + \alpha^{2}\cos^{2}\theta \right) .\eeq
On the slice $\bar{t}=0$ we have $v>0$ everywhere for $ |\alpha |<1$. In addition, since $\left( 1 + \alpha \cos\theta + \alpha^{2}\cos^{2}\theta \right)$ is always greater than $3/4$, for real $\alpha$ this hypersurface will become timelike near the horizon for $\alpha > \sqrt{7}/2 \sim 1.323$. 

On each slice with a constant value of $\bar{t}$, a two dimensional
marginally outer trapped surface (MOTS) can be searched for,
satisfying just $\theta_{(\ell)} =0$. In some spacetimes there may be
multiple MOTS on a given hypersurface. This will not occur in the
Vaidya spacetimes with linear mass function since the horizon
condition, $r=2m(v)$, is linear in $r$. The marginally outer trapped
surfaces can typically be stacked to form three dimensional trapping
horizons, provided the other conditions $\theta_{(n)} < 0$ and
$n^{a}\nabla_{a}\theta_{(n)} = 0$ are satisfied too (or a dynamical
horizon if the resulting surface is spacelike).  

For the case $\alpha=0$ the slicings will be spherically symmetric. The orbit of spherical rotations of each point will lie entirely in the hypersurface. This will lead to MOTS that are spherically symmetric, where each surface is just the orbit of spherical rotations. Commonly used slicings in numerical relativity such as constant mean curvature surfaces will typically be spherically symmetric in the Vaidya spacetime. In this case the horizons will be located uniquely at $r=2m(v)$ and every point on the surface will have the same value of the advanced time $v$. This will not be the case for surfaces with $\alpha \neq 0$. In these cases the mass parameter $m(v)$ will not necessarily take the same value on different points of the surface. The MOTS will then extend into regions where the matter is more compact and regions where it is less compact. Surfaces with $\alpha > 0$ will be the same as surfaces with $\alpha < 0$, with the north pole ($\theta=0$) interchanged with the south pole ($\theta=\pi$) or, equivalently $\cos\theta \rightarrow \cos(\pi-\theta)$.

Each different choice of $\bar{t}$ will lead to different MOTS than can be stacked to form different trapping horizons. Each of these trapping horizons will have an associated natural foliation, the one that makes each two dimensional surface a MOTS\@. However we can also consider the intersection of a given trapping horizon with a spacelike hypersurface of a given $\alpha$ value different from the natural one. We can compute the area of the these closed two spheres, although for $\alpha \neq 0$ they are not MOTS\@. For the spherically symmetric trapping horizon (SSTH) in Vaidya spacetime with linear mass function and $\dot{m} \ll 0$, the area will be
\beq A_\mathrm{SSTH} = 4\pi\left(2m\right)^{2}\left( 1 + 4\dot{m} + 2\dot{m}\frac{\bar{t}}{m} + \frac{4}{3}\dot{m}\alpha^{2} \right) .\eeq
A similar calculation can be performed for the intersections of the event horizon with the constant $\bar{t}$ surfaces. Taking the position of the event horizon to be (\ref{rehsmallmdot}) and again the approximation $\dot{m} \ll 0$ we find
\beq A_\mathrm{EH} = 4\pi\left(2m\right)^{2}\left( 1 + 12\dot{m} + 2\dot{m}\frac{\bar{t}}{m} + \frac{4}{3}\dot{m}\alpha^{2} \right) .\eeq
If we compare the area of the event horizon on a constant $\bar{t}$ hypersurface with the area of the event horizon on a spherically symmetric surface with $\alpha = 0$ but the same constant value of $t$ we see that the area will be greater if $\dot{m} > 0$ and smaller if $\dot{m}<0$.

\subsection{Horizon deformations}

The choice of different foliations leads to different MOTS that can be deformed into each other. Using these deformations it is possible to examine how certain geometric properties of surfaces change with the deformation \cite{Andersson:2005gq}. For example, it is possible to see how the area element, $\epsilon$, changes as MOTS are deformed into one another.

We can search for MOTS on slices with different values of $\bar{t}$ but because these marginally trapped surfaces do not lie on the same hypersurface it is difficult to compare them. If they do not intersect (and typically they don't) it may or may not be possible to find a trapping horizon that connects both surfaces. We want to distinguish between marginally trapped surfaces that are evolutions of one another and marginally trapped surfaces that cannot be evolved into one another along a trapping horizon.

For a given choice of $\bar{t}$ but two different values of $\alpha$ the two slices thus defined will intersect when $z=0$. The slice $t=0$ with $\alpha=0$ does not lie entirely to the past or future of the slice $\bar{t}=0$ with $\alpha = 1$. However, the horizons on these two slices will typically not intersect. On the two dimensional surface where these two slices do intersect, $z=0$, the horizon for $\alpha = 0$ is found to lie outside the horizon with $\alpha \neq 0$.

We can define a vector field, $X$, on surfaces that can be used to define variations of this surface, $\delta_{X}$. Following \cite{Booth:2006bn} we can write this vector field on a marginally trapped surface in terms of the two null normals to the surface as
\beq \label{boothdeform} X^{a} = B\ell^{a} - Cn^{a} ,\eeq
where $B$ and $C$ are functions on the surface. For deformations of the spherically symmetric surface where $\triangle = 0$ we have
\beq \label{spherboothdeform} X^{a} = B\frac{\partial}{\partial t} +C\frac{\partial}{\partial r} \eeq
for the Eddington-Finkelstein coordinates $t$ and $r$. The variation of intrinsic geometrical scalar properties can be calculated using
\beq \delta_{X}\phi = {\cal{L}}_{X}\phi , \eeq
and for general tensors, by projecting the Lie derivative onto the surface
\beq \delta_{X}w_{ab} = q_{a}^{c}q_{b}^{d}{\cal{L}}_{X}w_{ab} .\eeq
The variation of the area element $\epsilon$ for example satisfies
\beq \delta_{X}\epsilon = - C\theta_{(n)}\epsilon .\eeq
Variations of extrinsic properties are a little harder to calculate. We are particularly interested in deformations of the form
\beq \delta_{X}\theta_{(\ell)} = 0 ,\eeq
since this will generate a class of marginally outer trapped surfaces. The calculations in \cite{Booth:2006bn} give
\beq \delta_{X}\theta_{(\ell)} = - \d^{2}C + 2\tilde{\omega}^{a}\d_{a}C - C\delta_{(n)}\theta_{(\ell)} + B\delta_{(\ell)}\theta_{(\ell)} ,\eeq
where $\d$ is the covariant derivative compatible with the intrinsic metric of the two-surface. Notice that for radial $\ell^{a}$ we have (\ref{NEC}) and thus
\beq \delta_{(\ell)}\theta_{(\ell)} =  -\frac{2\dot{m}}{ r^{2}} ,\eeq
since the shear is zero. Furthermore, axisymmetric variations from the spherically symmetric surface will satisfy
\beq \frac{\d^{2}C}{\d\theta^{2}} + \cot\theta\frac{\d C}{\d\theta} + C - 2\dot{m}B = 0 .\eeq
In the case where both $B$ and $C$ are constant we get the evolution
along the spherically symmetric trapping horizon. In
\cite{Booth:2006bn} it was assumed that the deformations were
{\it{$l$-oriented}} such that $B>0$ everywhere. 
But another solution of this equation is
\beq B = \frac{{k}\cos\theta}{2\dot{m}} ,\eeq
\beq \label{ccos} C = -k\cos\theta ,\eeq
where $k$ is a constant of integration. Since the variation in this case is dependent on $\theta$ it will not generate spherically symmetric marginal surfaces, and since $B=C=0$ for $\theta=\pi/2$ this will not generate evolution along a trapping horizon. The area change in this case is easily computed as
\beq \int\delta_{X}\epsilon = -\int C \theta_{(n)} \epsilon = \frac{2k}{r}\int\cos\theta\sin\theta\d\theta\d\phi = 0 .\eeq
In this case, at least, the area of the spherically symmetric surface
is extremal. For a general axisymmetric variation expanded as a
Fourier series, the area of the spherically symmetric surface is
extremal provided the coefficients $a_{(n)}$ of the $\cos(n\theta)$
terms in $C$ are zero for even $n$.

\section{Numerical results for axisymmetric slicings}

We now compare the different trapping horizons one obtains for the axisymmetric slicings given by
(\ref{eq:tvralphaz}) where we consider the spherically symmetric mass functions (\ref{linearmass}),
(\ref{tanhlogmass}). For the numerical calculation in the following we use, as in previous work
\cite{Schnetter:2005ea}, the Cactus computational toolkit
\cite{Goodale:2002, cactus} which is widely used in numerical
relativity, for example to perform binary black hole, neutron star, or
stellar core collapse simulations. For each leaf we write the
3+1-decomposed Vaidya metric onto the Cartesian grid of Cactus. In each 'time step' we write the
next leaf of the slicing onto the Cartesian grid. This mimics a numerical time evolution of an
initial Cauchy slice of a dynamical BH spacetime. To locate the trapping horizon at each 'instant
of time' we use the apparent horizon finder AHFinderDirect
\cite{Thornburg:1996, Thornburg:2003sf}. This way we
obtain the location and area of the trapping horizon as a function of time. In the following we
compare these functions for the different trapping horizons we detect on the axisymmetric slicings
from above.
\subsection{Location of tilted versus untilted horizons}

The MOTS lying on the hypersurfaces $\bar{t} = v- r - \alpha z = 0$ with different values of $\alpha$ do not intersect. An examination of the marginally outer trapped surface with $\alpha = 0.83$ shows that the points at the north pole $\theta = 0$, equator $\theta = \pi/2$ and south pole $\theta = \pi$ have the four-dimensional spacetime coordinates given in Table (\ref{tab:horcoords})

\begin{table}[h]
\caption{Coordinate location of various points on surface with $\bar{t}=0$, $\alpha = 0.83$ and $m(v) =1.0 + 0.01v$}
\begin{center}
\begin{tabular}{ | c | c | c | c | }
\hline
    surface point & r & v & t \\ \hline
    zmax (north pole) & 2.0454 & 3.7498 & 1.7044 \\
    xmax (equator) & 2.0386 & 2.0386 & 0.0000 \\
    zmin (south pole) & 2.0247 & 0.3374 & -1.6872 \\ \hline
  \end{tabular}
\end{center}
\label{tab:horcoords}
\end{table}

Trapping horizons can be formed by stacking surfaces found on different hypersurfaces with different values of $\bar{t}$ but the same value of $\alpha$. These trapping horizons, each with their own value of $\alpha$ will intersect one another. This can also been seen in Table (\ref{tab:horcoords}). The points on the equator lie inside the horizon located on the spherically symmetric slicing $t = v-r =0$, where every point on the horizon has $r=2.0408$ and $t=0$. The point at the north pole lies inside the spherically symmetric horizon on the slice $t=1.7044$, which lies at $r=2.0756$, and the point at the south pole lies outside the spherically symmetric horizon on the slice $t=-1.6872$, which lies at $r=2.0064$.

In terms of the deformations discussed above (\ref{spherboothdeform}) we see that to deform the spherically symmetric surface at $t=0$ we need $B>0$ and $C>0$ at the north pole and $B<0$ and $C<0$ at the south pole, where in both cases the size of $C$ is much smaller than $B$.

\begin{figure}[!h]
\centering
\includegraphics[width=0.98\columnwidth,clip]{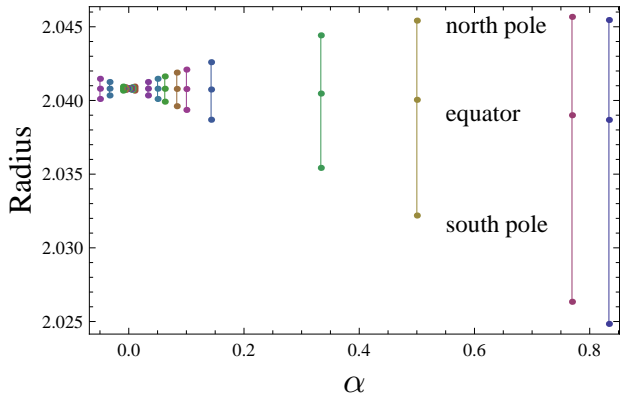}
\caption{The location ($r$-coordinate) of the north pole, equator and south pole versus different values of $\alpha$ for a linear mass function $m(v) = 1.0 + 0.01v$. The positions for negative $\alpha$ are the same as for positive $\alpha$ with the north pole and south pole interchanged. Near $\alpha=0$ there is no change in the position of the equator and the change in the position of the north pole and south pole are equal but opposite in sign.}
\label{fig:polepositions}
\end{figure}

Figure (\ref{fig:polepositions}) shows the locations ($r$ coordinate) of the north pole, equator and south poles for MOTS against $\alpha$ for a mass function of the form $m(v) = 1.0 + 0.01v$. For values of $\alpha$ less than zero the location of the MOTS is the same with the north pole $\theta=0$ and south pole $\theta=\pi$ interchanged. Near the spherically symmetric MOTS at $\alpha = 0$ the change in the $r$ coordinate of the equator is zero, whereas the change in the $r$ coordinates of the north and south poles is equal but opposite in sign. This supports the conjecture that for these cases $C$ is of the form (\ref{ccos}).

\subsection{Area dependence on $\alpha$}

\begin{figure}[!h]
\centering

\includegraphics[width=0.98\columnwidth,clip]{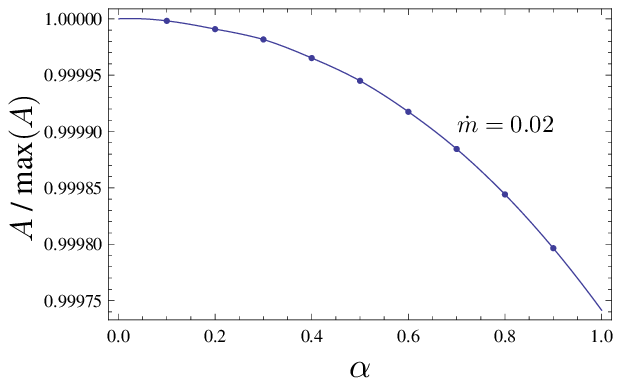}\\
\includegraphics[width=0.98\columnwidth,clip]{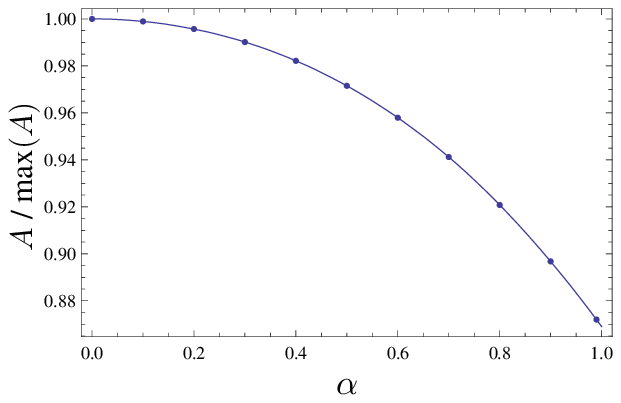}

\caption{Slicing dependence of the area at $\bar{t}$ for the linear mass function (top) with $m=1.0$ and $\dot{m}=0.02$
    and for the tanh log mass function (bottom) with $m=1.0$ and $T=1.0$}
%
%
\label{fig:alphaarea_linmass}
\label{fig:alphaarea_tanhlog}
\end{figure}

\begin{figure}[h]
\centering
\includegraphics[width=\columnwidth,clip]{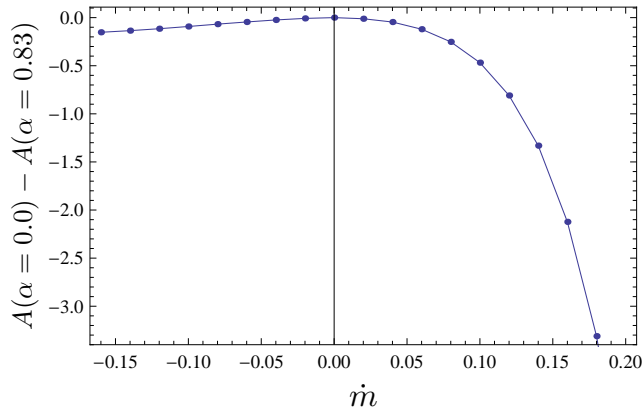}
\caption{Area difference for two slices in different Vaidya spacetimes with linear mass function and $m=1.0$, The difference between the areas increases for large $\dot{m}$. The untilted areas are larger than the tilted areas for $\dot{m}$ both positive and negative. However, the difference is much larger for $\dot{m}>0$. Only for $\dot{m}=0$ do the areas coincide.}
\label{fig:massdotarea_linmass}
\end{figure}

The results for the area dependence of $\alpha$ are shown in
Fig.(\ref{fig:alphaarea_linmass}) for the linear mass case and
Fig.(\ref{fig:alphaarea_tanhlog}) for the tanh log mass function. The
$\alpha=0$ surfaces have the largest area. As expected the area of the
tilted horizons is unchanged under a change in sign of $\alpha$. The
difference between the area on $\alpha = 0$ slices and $\alpha \neq 0$
slices depends on $\dot{m}$ as shown in
Fig.(\ref{fig:massdotarea_linmass}) for the linear mass function. It
is zero for $\dot{m}=0$, where the spacetime just reduces to the
Schwarzschild solution and the horizon is an isolated horizon. The
difference between the area of tilted and untilted horizons increases
with increasing magnitude of $\dot{m}$. The difference is slightly
larger for positive values of $\dot{m}$ than for negative values. This
difference in area does not seem to follow a simple power law and it
is difficult to extrapolate what the difference in areas will be for
values of $\dot{m}$ different form the values examined. However, the
difference in areas for values of $\dot{m}$ much lower than the lowest
positive value of $\dot{m}$ considered here ($\dot{m}=0.02$) are
likely to be much less than the $\sim 0.035\%$ difference found here
for that case.

\begin{figure}[h]
\centering
\includegraphics[width=0.98\columnwidth,clip]{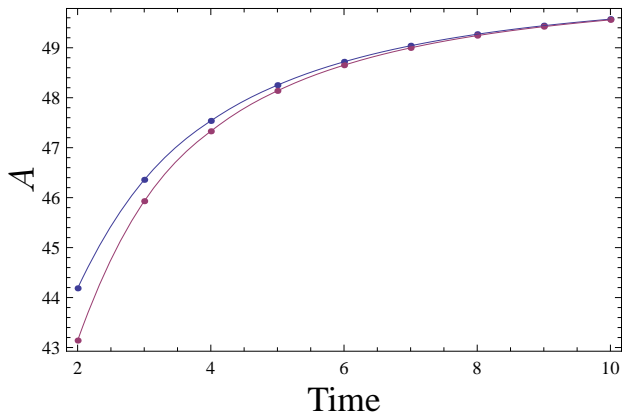}
\caption{Area versus time $\bar{t}$ for $\alpha = 0 $ and $\alpha = 0.83$ for the $\tanh\log$ mass function with $m=1.0$ and $T=1.0$. Both areas converge rapidly to the asymptotic isolated horizon area $16\pi$. The tilted area grows faster since it starts from a lower point.}
\label{fig:areatime_tanhlogmass}
\end{figure}

\begin{figure}
  \includegraphics[width=0.98\columnwidth,clip]{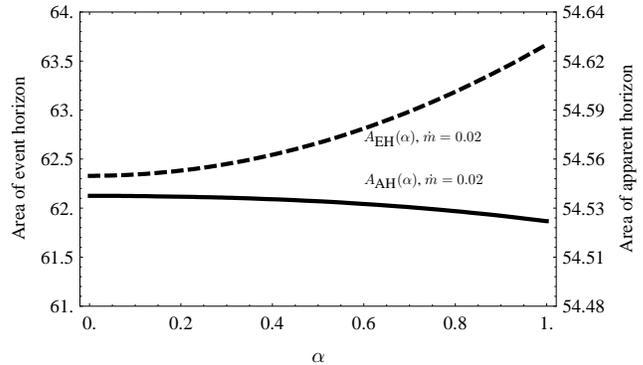}
  \caption{Area of the event horizon $A_{\mathrm{EH}}$ (dashed line) and the
    MOTS $A_{\mathrm{AH}}$ for different values of $\alpha$ for the linear
    mass function with $\dot{m} = 0.02$.  The y-axis on the left
    refers to the area of the event horizon and the axis on the right
    to the MOTS; the scale of $A_{\mathrm{AH}}$ has been expanded to better
    show its variation. The variation of $A_{\mathrm{EH}}$ is much
    larger than the variation of $A_{\mathrm{AH}}$ and the variation
    of $A_{\mathrm{AH}}$ is small relative to the difference between $A_{\mathrm{EH}}$ and $A_{\mathrm{AH}}$.}
  \label{fig:eh_ah_area}
\end{figure}

Finally, Fig.~\ref{fig:eh_ah_area} compares the area of the event
horizon (or more precisely, the intersection of the event horizon with
the spatial slices corresponding to different values of $\alpha$) with
the area of the MOTS as functions of $\alpha$.  It is interesting that
the variation in the event horizon area is much larger than for the
apparent horizon.  

\subsection{Dependence of rate of change of area on $\alpha$}

For the linear mass function the $\alpha\neq 0$ surfaces seem to have the same rate of change of area as the $\alpha = 0$ surfaces, at least to within numerical accuracy. For the tanh log mass function, as depicted in Fig.(\ref{fig:areatime_tanhlogmass}), the area of tilted horizons grows at a greater rate than the untilted areas and both converge to the asymptotic isolated horizon area, which will also be the asymptotic area of the event horizon. In this sense the spherically symmetric trapping horizon plays a special role as it is the one with the smallest rate of increase of area. This would seem to contradict the conjecture made in \cite{Gourgoulhon:2006uc} that the preferred trapping horizon should be the one where the area increases the most and approaches the event horizon the fastest. However, it is not known whether this behaviour is repeated in more general situations with other mass functions or other slicing conditions.

\section{Conclusion and discussion}

We have investigated the slicing dependence of the apparent horizons
in the Vaidya spacetime. For a given slice we have looked at three
types of horizons; the MOTS that lie on the slice, the intersection of
the slice with the event horizon and the intersection of the slice
with the spherically symmetric trapping horizon. We have examined
rather simple axisymmetric slicings in this simple spacetime. These
simple slicings do not exhaust all possible axisymmetric slicings and
there are many other possible slicings that are not axisymmetric.

Explicitly the location of the horizons is different for different
choices of horizon and different choices of slicing. However, for
slowly evolving horizons, the areas do not vary by much, although the
larger the rate at which the black hole accretes matter the larger the
difference in the areas. For the $\dot{m}=0.02$ Vaidya solution with
linear mass function the area can vary by approximately $0.03\%$ as
$\alpha$ is varied from $0$ to $1$. Beyond $\alpha = 1$ the Cauchy
surface becomes timelike for certain values of $\theta$. This is the
smallest value of $\dot{m}$ that we investigated but it is still much
larger than the mass accretion rate expected for astrophysical black
holes.

From a purely practical point of view we've found evidence that the
parameters of the black hole, derived from its trapping horizon, do
not change significantly when looking at reasonably simple
slicings. It is still unknown whether the properties would change
drastically for certain unusual slicings, but these are unlikely to
occur in normal numerical simulations. It is also reassuring that
these differences are much smaller than the parameter biases of up to
$10\%$ found between different post-Newtonian models
\cite{Buonanno:2009zt}. This variation of the parameters with foliation
will also occur if one uses the event horizon.

From a conceptual point of view, we have demonstrated explicitly the
known result that the location of the black hole surface and some of
its properties such as the area depend on the choice of the spacetime
slicing.  There are a variety of different possible responses to the
issue of non-unique\-ness of the trapping horizons in a given
spacetime.  The oldest approach is to focus purely on the event
horizon as the unique indicator of the black hole and its properties.
The location of the event horizon is independent of the choice of
foliation.  This approach however causes trouble in certain quantum
inspired spacetimes \cite{Hayward:2005gi,Nielsen:2008cr} where no
event horizon exists.  In numerical settings, the acausal definition
of the event horizon means that it is only known after a simulation is
finished, preventing it from being used to analyse the simulation's
state as it proceeds.

Another approach is to accept all horizons on an equal footing as
purely a property of the geometry. In this picture it is not clear how
to associate unique properties to the black hole such as a horizon
area or horizon angular momentum. Because the horizons intersect one
cannot rely on specifying, for example, the outermost trapping
horizon. However, it may be possible to formulate a generalized second
law for each possible foliation of spacetime and in this context use
the horizon defined by the chosen foliation.

A third approach is to look for the boundary of the region that admits trapped surfaces. This surface should be spherically symmetric in a spherically symmetric spacetime. If this surface lies strictly outside the spherically symmetric trapping horizon then, by the results of \cite{Ashtekar:2005ez}, it cannot itself admit the structure of a dynamical horizon. This surface will have a location that is independent of any given foliation, although it may not have a simple thermodynamic interpretation.

A fourth approach is to look for properties that select out a certain preferred trapping horizon in the spacetime \cite{Hayward:2009kr}, such as the spherically symmetric horizon in a spherically symmetric spacetime. This would require some additional condition to be imposed that selects out a unique trapping horizon from the many that occur in dynamical black hole spacetimes.

Despite their quasi-local nature, closed marginally trapped surfaces do have some non-local behaviour. Their dependence on the slicing is but one manifestation of this behaviour. The fact that marginally outer trapped surfaces can be found all the way to the event horizon \cite{BenDov:2006vw} is another manifestation. Although the event horizon is a fully non-local teleologically defined surface, it still acts as the boundary of outer trapped surfaces in a Vaidya spacetime that satisfies the null energy condition and asymptotes to a static Schwarzschild solution in the far future.

This behaviour is related to both the choice of the surface null
normals and the requirement that the marginally trapped surface be
closed.  Closed trapped surface cannot be found entirely in a flat
spacetime.  However, parts of a closed marginally trapped surface can
be found passing through a region of flat space where both null expansions are negative
\cite{Bengtsson:2008jr}.

It may be that in the quantum context only an effective horizon is meaningful or that the existence of many intersecting horizons contribute to the full path integral. For purely classical situations one may be able to refine the definition of the surface of a black hole.

\section{Appendix - some useful relations in the Vaidya spacetime}

These relations are provided here as a repository for reference in the main text. In the advanced null Eddington-Finkelstein coordinates $\left( v,r,\theta,\phi\right )$ the metric takes the form
\beq \d s^{2} = - \triangle \d v^{2} + 2\d r\d v + r^{2}\d\Omega^{2} ,\eeq
where $\triangle = 1- \frac{2m(v)}{r}$, $\triangle ' = \frac{2m(v)}{r^{2}}$, $\dot{\triangle} = -\frac{2\dot{m}}{r}$ and $\dot{m} = \partial_{v}m(v)$.\bigskip

\noindent Radial null vectors, with canonical normalization $\ell^{a}n_{a} = -1 $ and $n_{a} = -\partial_{v}r$:
\bea \ell^{a} &=& \left( 1,\frac{\triangle}{2},0,0\right) \,,\\
 \ell_{a} &=& \left(-\frac{\triangle}{2},1,0,0\right) \,,\\
 n^{a} &=& \left(0,-1,0,0\right) \,,\\
 n_{a} &=& \left(-1,0,0,0\right) \,.\eea
Expansions of the null normals:
\beq \theta_{(\ell)} = \frac{\triangle}{r}\,, \qquad \theta_{(n)} = -\frac{2}{r} \,.\eeq
The location of the spherically symmetric MOTS is obtained by setting
$\theta_{(\ell)} = 0$ leading to $r=2m(v)$.  The variations of the
expansions along $n^a$:
\bea n^{a}\nabla_{a}\theta_{(\ell)} &=&
\frac{\triangle}{r^{2}}-\frac{\triangle '}{r} \,,\\
\ell^{a}\nabla_{a}\theta_{(\ell)} &=& \frac{\dot{\triangle}}{r} - \frac{\triangle^{2}}{2r^{2}} +\frac{\triangle\triangle'}{2r}\,, \\
 n^{a}\nabla_{a}\theta_{(n)} &=& -\frac{2}{r^{2}} \,, \\
 \ell^{a}\nabla_{a}\theta_{(n)} &=& \frac{\triangle}{r^{2}} \,.\eea
Other optical scalars:
\beq \sigma_{(\ell)} = \sigma_{(n)} = \omega_{(\ell)} = \omega_{(n)} = 0 \,.\eeq
Surface gravities:
\bea \kappa_{(\ell)} &=& - n^{a}\ell^{b}\nabla_{b}\ell_{a} = \frac{\triangle '}{2}\,,\\
 \kappa_{(n)} &=& - \ell^{a}n^{b}\nabla_{b}n_{a} = 0 \,.\eea
Components of the energy-momentum tensor:
\bea T_{ab} &=& \frac{\dot{m}}{4\pi r^{2}}n_{a}n_{b}\,,\\ 
T_{ab}\ell^{a}\ell^{b} &=& \frac{\dot{m}}{4\pi r^{2}}\,,\\
 T_{ab}n^{a}\ell^{b} &=& T_{ab}n^{a}n^{b} = 0 \,.\eea
Normal to the trapping horizon $\triangle = 0$:
\beq N^{a} = \left( 1, -2\dot{m},0,0 \right) \,,\eeq
\beq N_{a} = \left(-2\dot{m}, 1,0,0 \right) \,.\eeq
Radial tangent to the trapping horizon
\beq V^{a} = \left(1, 2\dot{m},0,0\right) \,,\eeq
\beq V_{a} = \left(2\dot{m},1,0,0 \right) \,.\eeq
Variation of the area along the trapping horizon
\beq V^{a}\nabla_{a}A = \frac{4A\dot{m}}{r} \,.\eeq
Connection on the normal cotangent bundle
\beq \tilde{\omega}_{a} = - \tilde{q}^{b}_{a}n_{c}\nabla_{b}\ell^{c} = \left(0,0,0,0\right) \,.\eeq
Scalar curvature of the horizon
\beq \tilde{R} = \frac{2}{r^{2}} \,.\eeq
The null energy condition is satisfied if $E>0$. This is easy to see in a fiducial spherically symmetric null tetrad $\ell^{a}, n^{a}, m^{a}, \bar{m}^{a}$ since for a general null vector $v^{a}$ with
\beq \label{generalnull} v^{a} = A\ell^{a} + Bn^{a} + Cm^{a} + \bar{C}\bar{m}^{a} ,\eeq
then $T_{ab}v^{a}v^{b} = A^{2}T_{ab}\ell^{a}\ell^{b} = \frac{\dot{m}}{4\pi r^{2}}A^2(\ell^{a}n_{a})^{2}$. The null energy condition is closely related to whether there is positive or negative energy flowing into the black hole.\bigskip

\noindent If $\dot{m}\neq 0$ there are no isolated horizons in the Vaidya spacetime since the norm of the generalized null vector (\ref{generalnull}) is
\beq v^{a}v_{a} = -2AB + 2C\bar{C} = 0 .\eeq
For this null vector to be the generator of an isolated horizon it is required to satisfy $T_{ab}v^{a}v^{b} = 0$ giving $A=0$ and therefore from above $C=0$. Thus any null vector generating an isolated horizon must be proportional to $\theta_{(n)}$ but from the appendix we see $\theta_{(n)} = -2/r$. This only vanishes at infinity. There may be a degenerate planar isolated horizon ``at'' infinity.

\section{Acknowledgments} Alex Nielsen gratefully acknowledges
financial support from the Alexander von Humboldt Foundation and
hospitality at the Albert Einstein Institute;
Erik Schnetter acknowledges support from the NSF awards 0701566,
0721915, and 0904015.
The numerical solutions
were found using the AHFinderDirect thorn
\cite{Thornburg:1996,Thornburg:2003sf} of the Cactus framework
\cite{Goodale:2002, cactus} as part of the Einstein Toolkit \cite{einstein}.
We also thank Alberto Sesana and Jose Luis Jaramillo for helpful discussions.

During the work for this paper we were saddened to hear of the passing of Professor P. C. Vaidya, upon whose solution this work is based.

\end{document}